\documentclass[onecolumn,showpacs,preprintnumbers,amsmath,amssymb,a4paper,superscriptaddress]{revtex4}

\usepackage{graphicx}    
\usepackage{subfigure}   
\usepackage{citesort}    

\graphicspath{{Figures/},{./}}

\begin{document}


\title{Fear and its implications for stock markets}

\author{Ingve Simonsen}
\email{Ingve.Simonsen@phys.ntnu.no}
\affiliation{Department of physics, Norwegian University of Science
  and Technology (NTNU), NO-7491 Trondheim, Norway}
\affiliation{NORDITA, Blegdamsvej 17, DK-2100 Copenhagen {\O}, Denmark}

\author{Peter Toke Heden Ahlgren}
\email{peterahlgren@gmail.com}
\affiliation{The Niels Bohr Institute, Blegdamsvej 17, DK-2100
  Copenhagen, Denmark}

\author{Mogens H. Jensen}
\email{mhjensen@nbi.dk}
\affiliation{The Niels Bohr Institute, Blegdamsvej 17, DK-2100
  Copenhagen, Denmark}

\author{Raul Donangelo}
\email{donangel@if.ufrj.br}
\affiliation{Instituto de Fisica da UFRJ, Caixa Postal 68528,
  21941-972 Rio de Janeiro, Brazil}

\author{Kim Sneppen}
\email{sneppen@nbi.dk}
\affiliation{The Niels Bohr Institute, Blegdamsvej 17, DK-2100
  Copenhagen, Denmark}

\date{\today}

\begin{abstract}
  The value of stocks, indices and other assets, are examples of
  stochastic processes with unpredictable dynamics. In this paper, we
  discuss asymmetries in short term price movements that can not be
  associated with a long term positive trend. These empirical
  asymmetries predict that stock index drops are more common on a
  relatively short time scale than the corresponding raises. We
  present several empirical examples of such asymmetries. Furthermore,
  a simple model featuring occasional short periods of synchronized
  dropping prices for all stocks constituting the index is introduced
  with the aim of explaining these facts. The collective negative
  price movements are imagined triggered by external factors in our
  society, as well as internal to the economy, that create fear of the
  future among investors. This is parameterized by a ``fear factor''
  defining the frequency of synchronized events. It is demonstrated
  that such a simple fear factor model can reproduce several empirical
  facts concerning index asymmetries. It is also pointed out that in
  its simplest form, the model has certain shortcomings.
\end{abstract}

\maketitle


\section {Introduction and Motivation}

Extreme events such as the September 11, 2001 attack on New York city are
known to trigger rather systematically collapses in most sectors of
the economy. This does not happen because fundamental factors in the
economic system have worsened as a whole (from one day to another),
but because the prospects of the immediate future are considered
highly unknown. Investors simply {\em fear} for the future
consequences of such dramatic events, which is reflected in
dropping share prices. In other words, the share prices of a large
fraction of stocks show {\em collectively} a negative development
shortly after such a major triggering event~\cite{AsymPaper}.

These facts are rather well known, and one may give several other
similar examples. Fortunately, such extreme events
are not very frequent. One should therefore expect that collective
draw-downs are rare. On the contrary, we find that they are much more
frequent than one would have anticipated. One may say there is a
sequence of ``mini-crashes'' characterized by synchronized downward
asset price movements. As a consequence it is consistently more
probable --- up to a well defined (short) timescale --- to loose a
certain percentage of an investment placed in indices, than gaining
the same amount over the same time interval. This is what we call a
gain-loss asymmetry and it has been observed in different stock
indices including the Dow Jones Industrial Average~(DJIA), the SP500
and NASDAQ~\cite{Johansen}, but not, for instance, in foreign exchange
data~\cite{invfx}.

In this paper we will briefly revisit some of the empirical facts
of the gain-loss asymmetry. Then we suggest an explanation of the
phenomenon by introducing a simple (fear factor) model. The model
incorporates the concept of a synchronized event, among the
otherwise uncorrelated stocks that compose the index. This effect
is seen as a consequence of risk aversion (or fear for the future)
among the investors triggered by factors external, as well as
internal, to the market. In our interpretation, the results show
that the concept of fear has a deeper and more profound
consequence on the dynamics of the stock market than one might
have initially anticipated.

\section{The inverse statistics approach}

A new statistical method, known as {\em inverse statistics} has
recently been introduced~\cite{Mogens,optihori,gainloss,invfx}. In
economics, it represents a {\em time-dependent} measure of the
performance of an asset.  Let $S(t)$ denote the asset price at time
$t$. The logarithmic return at time $t$, calculated over a time
interval $\Delta t$, is defined
as~\cite{Book:Bouchaud-2000,Book:Mantegna-2000,Book:Hull-2000}
$r_{\Delta t}(t) = s(t+\Delta t)- s(t)$, where $s(t) = \ln S(t)$.  We
consider a situation in which an investor aims at a given return
level, $\rho$, that may be positive (being ``long'' on the market) or
negative (being ``short'' on the market). If the investment is made at
time $t$ the inverse statistics, also known as the {\em investment
  horizon}, is defined as the {\em shortest} time interval
$\tau_\rho(t)=\Delta t$ fulfilling the inequality $r_{\Delta t}(t)\geq
\rho$ when $\rho\geq 0$, or $r_{\Delta t}(t)\leq \rho$ when $\rho<0$.

The inverse statistics histogram, or in economics, the investment
horizon distribution, $p\left( \tau_\rho\right)$, is the distribution
of all available waiting times $\tau_\rho(t)$ obtained by moving
through time $t$ of the time series to be analyzed
(Fig.~\ref{Fig:DJIA_PDF}(a)). Notice that these methods, unlike the
return distribution approach, do {\em not} require that data are
equidistantly sampled. It is therefore well suited for tick-to-tick
data.

If the return level $\rho$ is not too small the distribution $p\left(
  \tau_\rho\right)$ has a well defined maximum, see
Fig.~\ref{Fig:DJIA_PDF}(a). This occurs because it takes time to drive
prices through a certain level. The most probable (waiting) time, {\it
  i.e.} the maximum of the distribution, corresponds to what has been
termed the {\em optimal investment horizon}~\cite{optihori} for a
given return level, $\rho$, and will be denoted $\tau_\rho^*$ below.

\section{Empirical results}
\label{sec:empiricalfindings}

In this section, we present some empirical results on the inverse
statistics. The data set used is the daily close of
the DJIA covering its entire history from 1896 till today.
Fig.~\ref{Fig:DJIA_PDF}(a) depicts the empirical inverse statistics
histograms --- the investment horizon distribution --- for
(logarithmic) return levels of $\rho=0.05$ (open blue circles) and
$\rho=-0.05$ (open red squares). The histograms possess well defined
and pronounced maxima, the {\em optimal investment horizons}, followed
by long $1/t^{3/2}$ power-law tails that are well
understood~\cite{Book:Bouchaud-2000,Book:Mantegna-2000,Book:Hull-2000,Book:Johnson-2003}.
The solid lines in Fig.~\ref{Fig:DJIA_PDF}(a) represent generalized
inverse Gamma distributions~\cite{optihori} fitted towards the
empirical histograms.  This particular functional form is a natural
candidate since it can be shown that the investment horizon
distribution is an inverse Gamma distribution \footnote{In mathematics,
  this particular distribution is also known as the L\'evy
  distribution in honor of the French mathematician Paul Pierre
  L\'evy. In physics, however, a general class of (stable) fat-tailed
  distributions is usually ´called by this name.}, $p(x)\sim
{\rm exp}(a/2x)/x^{3/2}$ ($a$ being a parameter), if the analyzed asset
price process is (a pure) geometrical Brownian
motion~\cite{Book:Karlin-1966,Book:Redner-2001}.

A striking feature of Fig.~\ref{Fig:DJIA_PDF}(a) is that the optimal
investment horizons with equivalent magnitude of return level, but
opposite signs, are {\em different}. Thus the market as a whole,
monitored by the DJIA, exhibits a fundamental {\em gain-loss
  asymmetry}.  As mentioned above other indices, such as SP500 and
NASDAQ, also show this asymmetry~\cite{Johansen}, while, for instance,
foreign exchange data do not~\cite{invfx}.

It is even more surprising that a similar well-pronounced asymmetry is
{\em not} found for any of the individual stocks constituting the
DJIA~\cite{Johansen}.  This can be observed from the insert of
Fig.~\ref{Fig:DJIA_PDF}(a), which shows the results of applying the
same procedure, individually, to these stocks, and subsequently
averaging to improve statistics.

Fig.~\ref{Fig:DJIA_PDF_scaling}(a) depicts the empirical dependence of
the optimal investment horizon (the maximum of the distribution),
$\tau_\rho^*$, as a function of the return level $\rho$. If the
underlying stochastic price process is a geometrical Brownian motion,
then one can show that $\tau_\rho^*\sim \left|\rho\right|^\gamma$,
with $\gamma=2$, valid for all return levels $\rho$ (indicated by the
lower dashed line in Fig.~\ref{Fig:DJIA_PDF_scaling}(a)). Instead one
empirically observes a different behavior with a weak ($\gamma\approx
0$), or no, dependence on the return level when it is {\em small}
compared to the (daily) volatility, $\sigma$, of the index. For
instance the DJIA daily volatility is about $\sigma_{\tiny
  DJIA}\approx 1$\%. On the other hand a cross-over can be observed,
for values of $\rho$ somewhat larger than $\sigma$, to a regime where
the exponent $\gamma$ is in the range of $1.8$--$2$.  Based on the
empirical findings we do not insist on a power-law dependence of
$\tau_\rho^*$ on $\rho$. The statistics is too poor to conclude on
this issue, and there seem to be even some $\rho$ dependence in
$\gamma$. Other groups though~\cite{ZhouYuan,Poland}, have found
indications of similar power-law behavior in both emerging and liquid
markets supporting our findings.  An additional interesting and
apparent feature to notice from Fig.~\ref{Fig:DJIA_PDF_scaling}(a) is
the consistent, almost constant, relative gain-loss asymmetry in a
significant wide range of return levels.

In light of these empirical findings the following interesting and
fundamental question arises: Why does the index exhibit a pronounced
asymmetry, while the individual stocks do not? This question
is addressed by the model introduced below.

\section{The fear factor model}
\label{sec:model}

Recently the present authors introduced a so-called {\em fear factor
  model} in order to explain the empirical gain-loss
asymmetry~\cite{AsymPaper}.  The main idea is the presence of
occasional, short periods of dropping stock prices synchronized
between all $N$ stocks contained in the stock index. In essence these
collective drops are the cause (in the model) of the asymmetry in the
index~\cite{AsymPaper}. We rationalize such behavior with the
emergence of anxiety and fear among investors. Since we are mainly
interested in day-to-day behavior of the market, it will be assumed
that the stochastic processes of the stocks are all equivalent and
consistent with a {\em geometrical Brownian
  motion}~\cite{Book:Bouchaud-2000,Book:Mantegna-2000,farmer}. This
implies that the logarithm of the stock prices, $s_i(t)=\ln S_i(t)$,
follow standard, unbiased, random walks
\begin{align}
  s_i(t+1) = s_i(t) + \varepsilon_i(t) \delta\;, \qquad i=1,\ldots,N,
\label{logprice}
\end{align}
where $\delta>0$ denotes the common fixed log-price increment (by
assumption), and $\varepsilon_i(t)=\pm 1$ is a random time-dependent
direction variable. At certain time steps, chosen randomly with fear
factor probability $p$, all stocks {\em synchronize} a collective draw
down ($\varepsilon_i=-1$). For the remaining time steps, the different
stocks move independently of one another. To assure that the overall
dynamics of every stock is behaving equivalent to a geometric Brownian
motion, a slight upward drift, quantified by the probability for a
stock to move up $q=1/(2(1-p))$~\cite{AsymPaper}, is introduced.
This ``compensating'' drift only governs the non-synchronized periods.
>From the price realizations of the $N$ single stocks, one may
construct the corresponding price-weighted index, like in the DJIA,
according to
\begin{align}
 I(t) = \frac{1}{d(t)} \sum_{i=1}^N S_i(t) \;=\;
          \frac{1}{d(t)} \sum_{i=1}^N \exp \{s_i(t)\}\;.
\label{index1}
\end{align}
Here $d(t)$ denotes the divisor of the index at time $t$ that for
simplicity has been fixed to the value $d(t)=N$.  Some consideration
is needed when choosing the value of $\delta$.  If
$\delta$ is too small the daily index volatility \footnote{Note that
  the index volatility does in principle depend on $\delta$, the
  number of stocks $N$, as well as the fear factor $p$.} will not be 
large enough to reach the return barrier $\rho$ within an appropriate
time.  On the other hand, when $\delta$ is too large it will cause a
crossing of the negative return barrier no later than the {\em first}
occurring synchronous step (if $\left|\rho\right|$ is not too large).
Under such circumstances, the optimal investment horizon for negative
returns ($\tau_{-\left|\rho\right|}^*$) will only to a very little
extent depend on $\rho$ which is inconsistent with empirical
observations.  Therefore, the parameter $\delta$ should be chosen
large enough, relative to $\left|\rho\right|$, to cause the asymmetry,
but not too large to dominate fully whenever a synchronous step
occurs.  A balanced two-state system is the working mechanism of the
model --- dominating calm behavior interrupted by short-lived bursts
of {\em fear}.  For more technical details on the model, the
interested reader is referred to Ref.~\cite{AsymPaper}.

It is important to realize that the asymmetric investment horizons
obtained with the model stems from the very simple synchronization
events between stocks isolated in time, and not by means of
higher-order correlations. The cause of these simultaneous
drawdowns {\it could} be both internal and external to the market,
but no such distinctions are needed to create the dynamics and
asymmetry of the fear factor model. This differs --- at least
model wise --- the inverse statistics asymmetry from the {\em
leverage} phenomenon reported in {\it
e.g.}~\cite{BouchaudLeverage} and elsewhere. Though the two
phenomena conceptually are related, the asymmetric gain-loss
horizons can be simulated without involving complex {\em
stochastic volatility} models.

The minimalism of the model also involves aspects that are not
entirely realistic. Work is in progress to extend the model by
including features of more realistic origin. In particular, we
have included splittings, mergers and
replacements as well as selected economic sectors which have their own
fear factors. Moreover, other extensions include more realistic
(fat-tailed) price increment
distributions~\cite{Book:Bouchaud-2000,Book:Mantegna-2000,Book:Hull-2000,mandelbrot62}
as well as time-dependent stochastic volatility for the single
stocks~\cite{engle,engle-patton,Book:Bouchaud-2000,Book:Mantegna-2000,Book:Hull-2000}.
The detailed results of these extensions will be reported
elsewhere~\cite{FurtherWork}.

\section{Results and discussion}

\begin{figure}
  \begin{center}
    \subfigure[The empirical results for the DJIA.]{\includegraphics*[width=0.7\columnwidth,height=0.5\columnwidth]{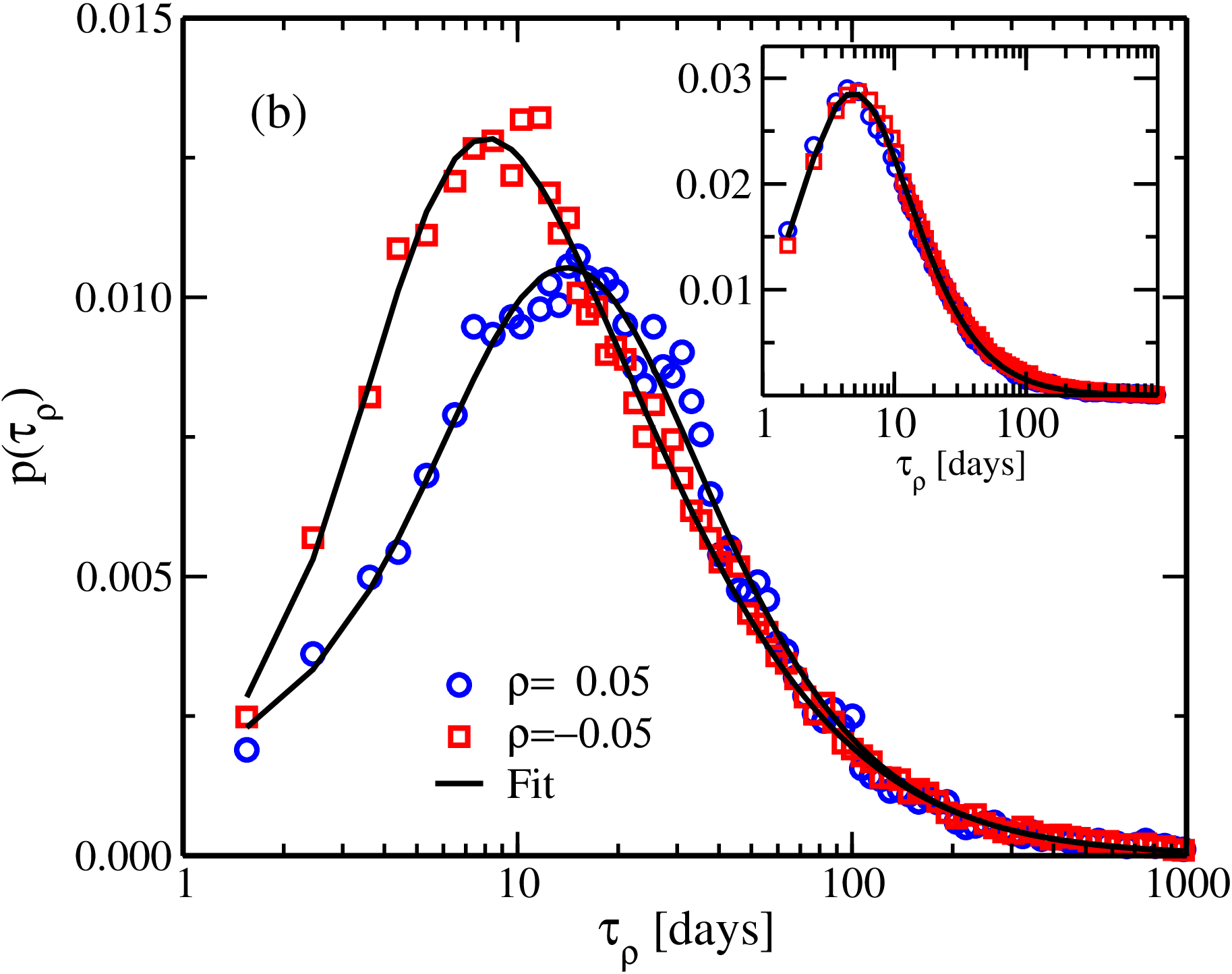}}\\*[2mm]
    \subfigure[The results for the fear factor model.]{\includegraphics*[width=0.7\columnwidth,height=0.5\columnwidth]{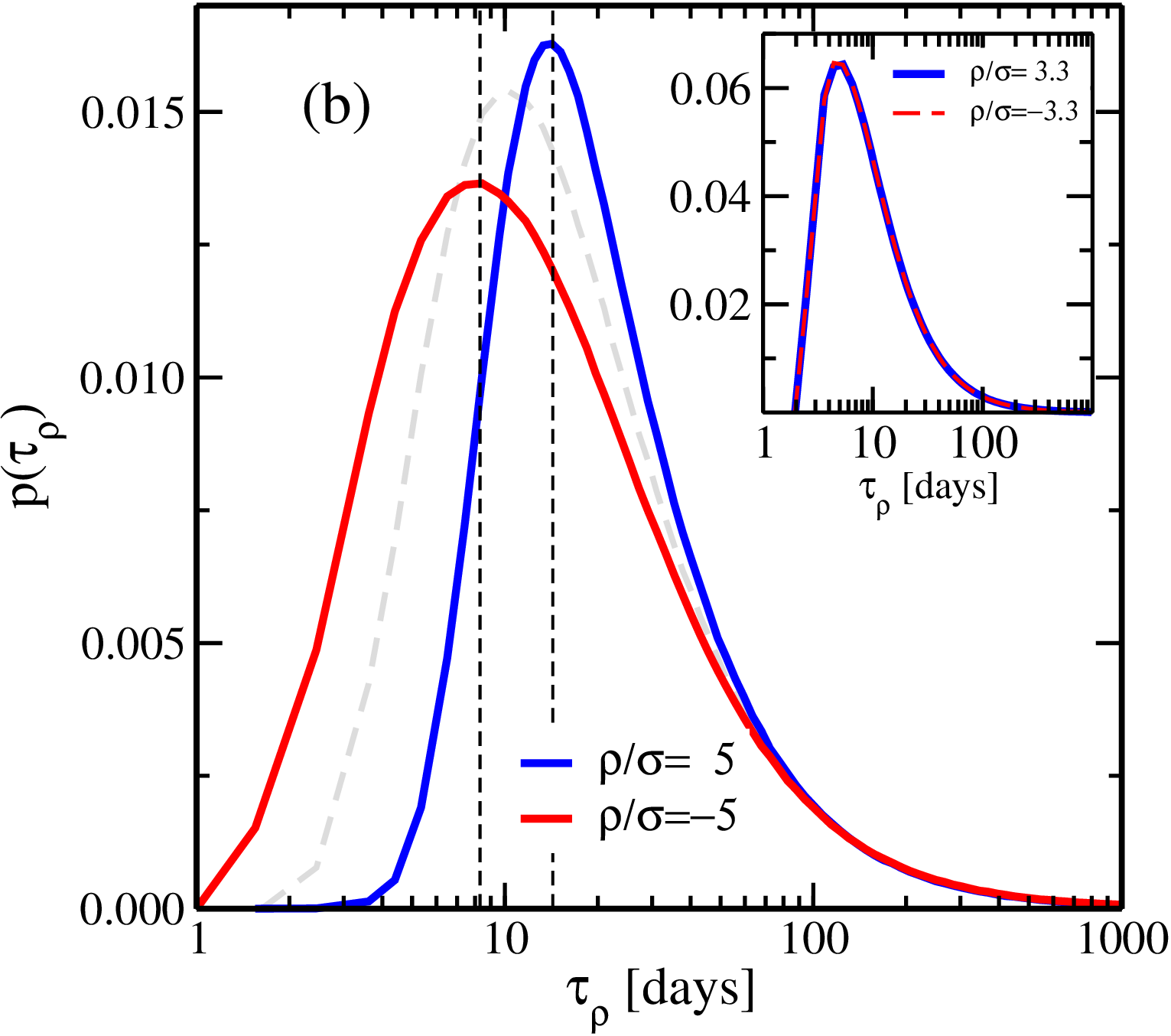}}
    \caption{(Color online) The inverse statistics distributions : (a)
      The panel shows histograms of the inverse statistics for the
      DJIA obtained on the basis of the empirical daily close data
      covering its entire history of 110 years from 1896.  The red
      open squares are obtained using a loss level of $\rho = - 0.05$
      and the blue open circles are obtained using a gain level of
      $\rho = + 0.05$ and both distributions are normalized.  Note the
      clear asymmetry between the loss and the gain statistics.  The
      full curves are regression fits using a generalized inverse
      Gamma distribution~\cite{optihori,gainloss,invfx}.  The inset shows
      the distributions obtained from using the same procedure on the
      individual stocks of the DJIA, and subsequently averaging over
      the stocks.  Notice that the asymmetry is absent for individual
      stocks. (b) The results for the inverse statistics obtained within the
      fear factor  model applying parameters characterizing the  DJIA and
      used to produce the empirical results of
      Fig.~\protect\ref{Fig:DJIA_PDF}(a). In particular it was used that
      the index consists of $N=30$ stocks and the return level was
      set to $\rho/\sigma=5$, where $\rho$ is the return level and $\sigma$
      denotes the daily volatility of the index.  In the model the
      index volatility, $\sigma$, should reflect the observed $1$\%
      daily volatility of the DJIA, and the $\rho/\sigma=\pm 5$ therefore
      corresponds to $\rho=\pm 5$\% in
      Fig.~\protect\ref{Fig:DJIA_PDF}(a).  A fear
      factor of $p=0.05$ was chosen to reproduce the positions of the
      two asymmetric maxima appearing in
      Fig.~\protect\ref{Fig:DJIA_PDF}(a) and
      indicated by dashed vertical lines.  The dashed thick line is
      the result for a fear-factor parameter $p=0$, in which case the
      asymmetry vanishes.  As in Fig.~\protect\ref{Fig:DJIA_PDF}(a), the inset
      shows the loss and gain distributions for the individual stocks
      in the model.  Notice, that here the asymmetry is also absent. }
    \label{Fig:DJIA_PDF}
  \end{center}
\end{figure}

We will now address the results that can be obtained by the fear
factor model and compare them with the empirical findings.
Fig.~\ref{Fig:DJIA_PDF}(b) shows that the model indeed produces a
clear gain-loss asymmetry in the inverse statistic histograms. Hence,
the main goal of the model is obtained. Moreover, the
investment horizon distributions are qualitatively very similar to
what is found empirically for the DJIA (cf.
Fig.~\ref{Fig:DJIA_PDF}(a)). In particular, one observes from
Fig.~\ref{Fig:DJIA_PDF}(b) that the positions of the peaks found
empirically (vertical dashed lines) are predicted rather accurately by
the model. To produce the results of Fig.~\ref{Fig:DJIA_PDF}(b), a
fear factor of $p=0.05$ was used. Furthermore it is observed, as
expected, that the model with $p=0$ does not produce any asymmetry
(grey dashed line in Fig.~\ref{Fig:DJIA_PDF}(b)).

A detailed comparison of the shapes of the empirical and the
modelled inverse statistics distribution curves 
reveal some minor differences,
especially regarding short waiting times and the height of the
$\rho>0$ histogram.  One could find simple explanations for these
differences, such as the fact that the model does not consider a
realistic jump size distribution, or even that it does not include
an ``optimism factor'' synchronizing draw-ups.  This would result
in a wider $\rho>0$ distribution for short waiting times, and
additionally would lower the value of the maximum.  Some of 
these shortcomings of the minimalistic model has already been
dealt with in Ref. ~\cite{FurtherWork}.
For the sake of this paper, however, none of these extension will
be further discussed in detail.

\begin{figure}[t]
  \begin{center}
    \leavevmode
   \subfigure[The empirical results for the DJIA.]{\includegraphics*[width=0.7\columnwidth,height=0.5\columnwidth]{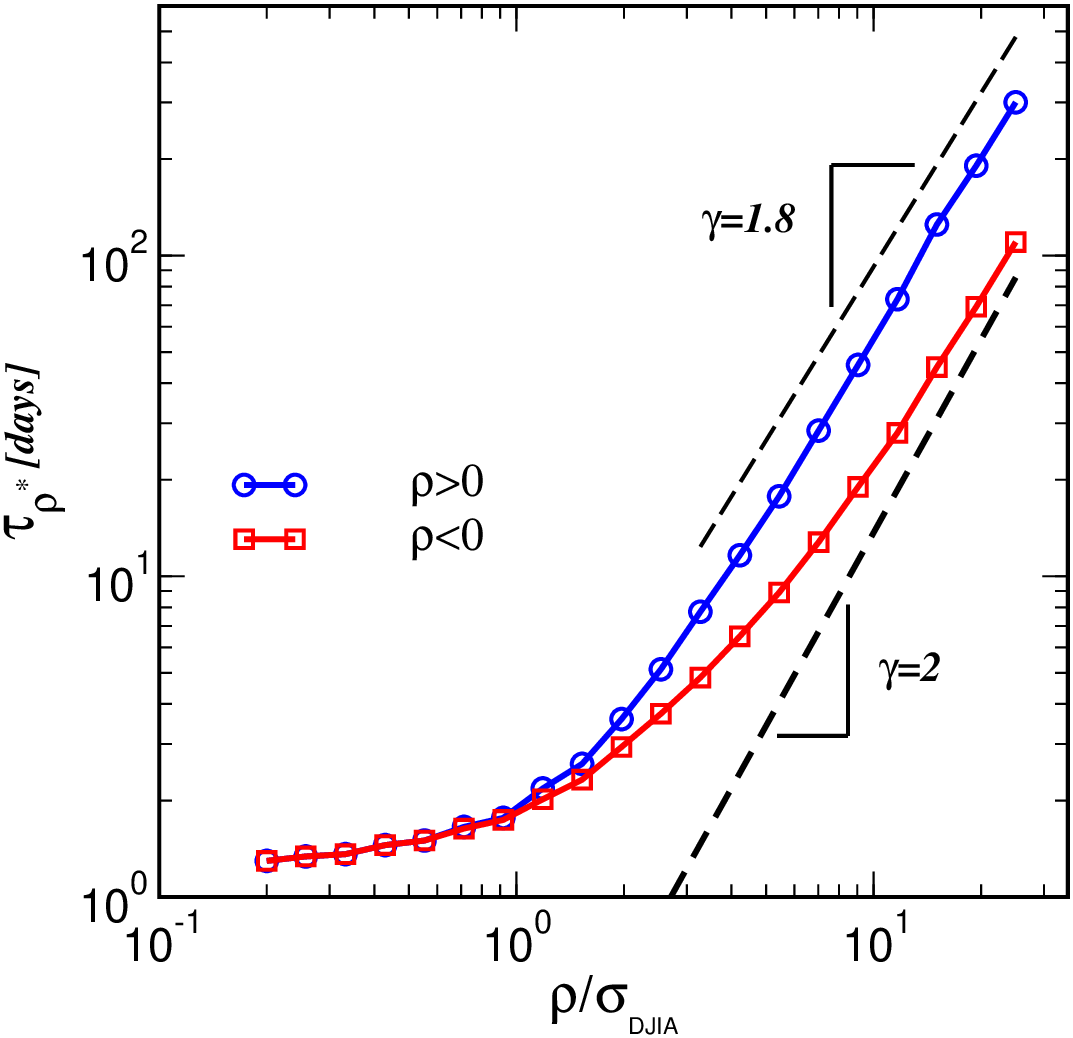}}\\*[2mm]
   \subfigure[The results for the fear factor model.]{\includegraphics*[width=0.7\columnwidth,height=0.5\columnwidth]{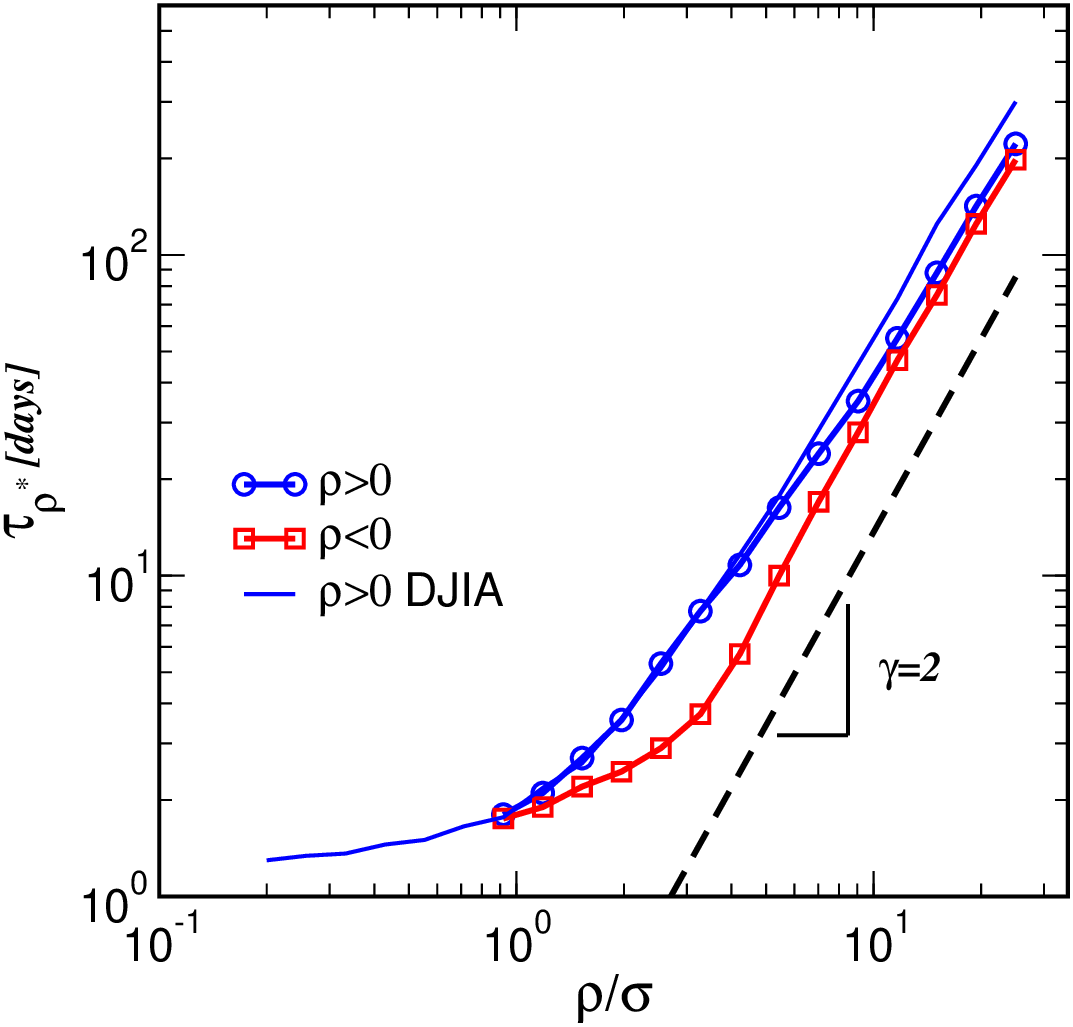}}
   \caption{The dependence of the optimal investment horizon on the
     return level (scaled with the daily volatility): (a) The DJIA
     empirical optimal investment horizon $\tau^*_\rho$ for
     positive~(open circles) and negative~(open squares) levels of
     return $\pm \rho$.  The daily volatility used for the rescaling
     was $\sigma_{\mbox{\tiny DJIA}}\approx 1\%$.  
     In the case where $\rho<0$ one has used $-\rho$ on the abscissa
     for reasons of comparison.  If a geometrical Brownian price
     process is assumed, one will have $\tau^*_\rho\sim \rho^\gamma$
     with $\gamma=2$ for all values of $\rho$. Such a scaling behavior
     is indicated by the lower dashed line in the graph.  Empirically
     one finds values of $\gamma\simeq 1.8$~(upper dashed line),
     but only for large values of the return. (b) Results of the fear
     factor model analog to the empirical DJIA results of
     Fig.~\protect\ref{Fig:DJIA_PDF_scaling}(a). The parameters used
     to produce the model results of this figure were those given in
     the caption of Fig.~\protect\ref{Fig:DJIA_PDF}.  }
    \label{Fig:DJIA_PDF_scaling}
  \end{center}
\end{figure}

Fig.~\ref{Fig:DJIA_PDF_scaling}(b) depicts the optimal investment
horizon {\it vs.} return level obtained from our fear factor
model. It is observed that for $\rho>0$ the empirical result for
the DJIA (solid line in Fig.~\ref{Fig:DJIA_PDF_scaling}(b)) is
reasonably well reproduced.  One exception is for the largest
return levels, where a value of $\gamma=2$ seems to be
asymptotically approached.  This might not be so unexpected since
this is the geometric Brownian motion value. However, the case is
different for $\rho<0$. Here the empirical behavior is not
reproduced that accurately.  Consistent with empirical findings, a
gain-loss asymmetry gap,
$\tau_{+\left|\rho\right|}^*-\tau_{-\left|\rho\right|}^*$, opens
up for return levels $\rho$ comparable to the volatility of the
index $\sigma$.  Unlike empirical observed results
(Fig.~\ref{Fig:DJIA_PDF_scaling}(a)), the gap, however, decreases
for larger return levels. The numerical data seems also to
indicate that the closing of the gain-loss asymmetry gap results
in a data collapse to a universal $\tau_\rho^*\sim \rho^\gamma$
curve with exponent $\gamma=2$.

We will now argue why this is a plausible scenario.  Even during a
synchronous event, when all stocks drop simultaneously, there is a
upper limit on the drop of the index value.  One can readily show
({\it cf.} Eq.~(4) of Ref.~\cite{AsymPaper}) that the relative returns
of the index during a synchronous event occurring at  $t+1$ is
\begin{align}
  \frac{\Delta I(t)}{I(t)}
         = \frac{I(t+1)-I(t)}{I(t)}
         = \exp(-\delta)- 1 <0,
\end{align}
which also is a good approximation to the corresponding logarithmic
return as long as $I(t)\gg \Delta I(t)$~\cite{Book:Mantegna-2000}.
This synchronous index drop sets a scale for the problem.  One has
essentially three different regions, all with different properties,
depending on the applied level of return. They are characterized by
the return level $\rho$ being {\it (i)} much smaller than; {\it (ii)}
comparable to; or {\it (iii)} much larger than the synchronous index
drop $\exp(-\delta)-1$. In case {\it (i)}, the synchronization does
not result in a pronounced effect, and there is essentially no
dependence on the return level or its sign. For the intermediate
range, where $\rho$ is comparable to $\exp(-\delta)-1$, the asymmetric
effect is pronounced since no equivalent positive returns are very
probable for the index (unless the fear factor is very small).
Specifically, whenever $\rho<\exp(-\delta)-1$ one collective draw-down
event is sufficient to cross the lower barrier of the index, thereby
resulting in an exit time coinciding with the time of the
synchronization. Of course this is not the case when $\rho>0$ giving
the working mechanism in the model for the asymmetry at short time
scales.  For the final case, where $\rho \gg \exp(-\delta)-1$, neither
the synchronized downward movements, or the sign of the return level,
play an important role for the barrier crossing. However, in contrast
to case {\it (i)} above, the waiting times are now much longer, so
that the geometrical Brownian character of the stock process is
observed.  This is reflected in Fig.~\ref{Fig:DJIA_PDF_scaling}(b) by
the collapse onto an apparent common scaling behavior with $\gamma=2$
independent of the sign of the return level.

\begin{figure}
\includegraphics*[width=0.7\columnwidth]{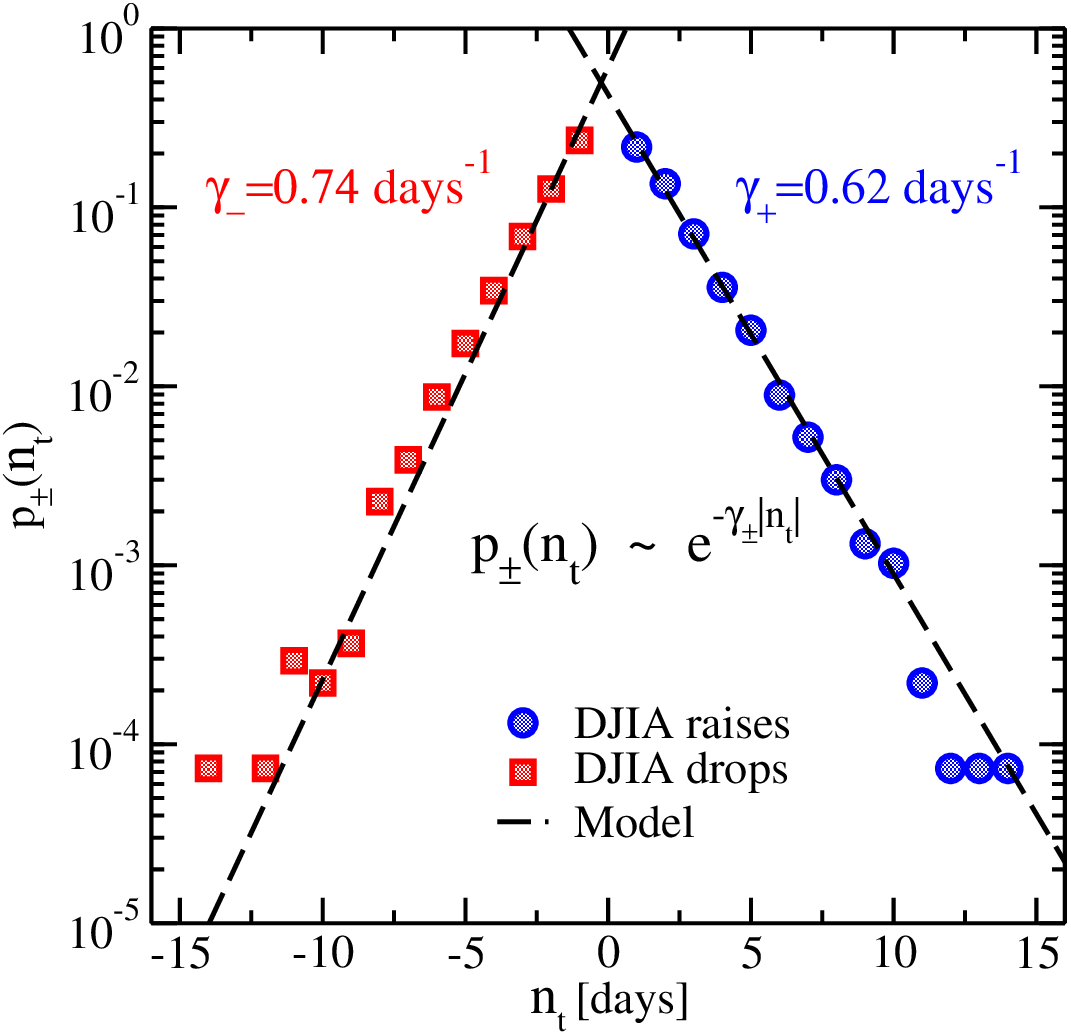}
\caption{(Color online) The distribution, $p_\pm(n_t)$, of the number
  of {\em consecutive} days in a row, $n_t$, the DJIA index is
  dropping (filled red squares) or raising (filled blue circles
  squares). We have adopted the convention that drops correspond to
  negative values of $n_t$ while raises correspond 
  to positive. The exponential
  rates $\gamma_\pm$ for the empirical DJIA data were
  determined by regression to
  $\gamma_+=(0.62\pm0.01)\,\mbox{days}^{-1}$ and
  $\gamma_-=(0.74\pm0.03)\,\mbox{days}^{-1}$.  The dashed lines
  correspond to the prediction of the fear factor model using the
  parameters of Fig.~\protect\ref{Fig:DJIA_PDF} except that the fear
  factor was lowered slightly ($p=0.02$). Notice that only one single
  parameter, the fear factor $p$, had to be adjusted in order to
  correctly reproduce the {\em two} empirical rates.}
\label{fig:Motifs}
\end{figure}

The last topic to be addressed in this paper is also related to an
asymmetry, but takes a somewhat different shape from what was
previously considered. By studying the probability that the DJIA
index decreases, respectively increases, from day to day, we have
found a $9\%$ larger probability for the index to decrease rather
than increase. This information led us to a more systematic study by
considering the number of {\em consecutive} time steps, $n_t$, the
index drops or raises. This probability distribution will be
denoted by $p_\pm(n_t)$, where the subscripts $+/-$ refers to
price raise/drop.  The open symbols of Fig.~\ref{fig:Motifs} show
that the empirical results, based on daily DJIA data, are
consistent with decaying exponentials of the form $p_\pm(n_t)\sim
\exp(-\gamma_\pm \left|n_t\right|)$, where $\gamma_\pm>0$ are
parameters (or rates). It is surprising to observe that also this
measure exhibits an asymmetry since $\gamma_+ \neq \gamma_-$.
These rates, obtained by exponential regression fits to the
empirical DJIA data, are $\gamma_+ =(0.62 \pm
0.01)\,\mbox{days}^{-1}$ and $\gamma_- =(0.74 \pm
0.03)\,\mbox{days}^{-1}$.  What does the fear factor model,
indicate for the same probabilities?  In Fig.~\ref{fig:Motifs} the
dashed lines are the predictions of the model and they reproduce
the empirical facts surprisingly well. They correspond to the
following parameters $\gamma_+=0.62 \,\mbox{days}^{-1}$ and
$\gamma_-=0.78\,\mbox{days}^{-1}$ for the raise and drop curves,
respectively.  However, the value of the fear factor necessary to
obtain these results was $p=0.02$. This is slightly lower than the
value giving consistent results for the inverse statistics
histograms of Figs.~\ref{Fig:DJIA_PDF_scaling}. In this respect,
the model has an obvious deficiency. It should be stressed,
though, that it is a highly non-trivial task, with one adjustable
parameter, to reproduce correctly the two different rates
($\gamma_\pm$) for the two probabilities. That such a good
quantitative agreement with real data is possible must be seen as
a strength of the presented model.

\section{Conclusions and outlooks}

In conclusion, we have briefly reviewed what seems to be a new
stylized fact for stock indices that show a pronounced gain-loss
asymmetry.  We have described a so-called minimalistic ``fear
factor'' model that conceptually attributes this phenomenon to
occasional synchronizations of the composing stocks during some
(short) time periods due to fear emerging spontaneously among
investors likely triggered by external world events. This
minimalistic model do represent a possible mechanism for the
gain-loss asymmetry, and it reproduces many of the empirical facts
of the inverse statistics.

\section*{Acknowledgements}

We are grateful for constructive comments from Ian Dodd and Anders
Johansen.  This research was supported in part by the ``Models of
Life'' Center under the Danish National Research Foundation. R.D.
acknowledges support from CNPq and FAPERJ (Brazil).



\newpage

\end{document}